\documentclass{emulateapj}

\usepackage{apjfonts}
\usepackage{natbib,graphicx}

\newcommand{\xmm}{{\it XMM-Newton}}
\newcommand{\chandra}{{\it Chandra}}
\newcommand{\swift}{{\it Swift}}
\newcommand{\rosat}{{\it ROSAT}}
\newcommand{\galex}{{\it Galex}}
\newcommand{\rxte}{{\it RXTE}}
\newcommand{\Ell}{{\cal L}}

\begin{document}

\title{V723\,Cassiopeia still on in X-rays: A bright Super Soft
Source 12 years after outburst}

\author{J.-U. Ness\altaffilmark{1}, G. Schwarz\altaffilmark{2},
S. Starrfield\altaffilmark{1}, J.P. Osborne\altaffilmark{3},
K.L. Page\altaffilmark{3}, A.P. Beardmore\altaffilmark{3},
R.M. Wagner\altaffilmark{4}, and C.E. Woodward\altaffilmark{5}}

\altaffiltext{1}{School of Earth and Space Exploration, Arizona
State University, Tempe, AZ 85287-1404, USA: Jan-Uwe.Ness}
\altaffiltext{2}{West Chester University of Pennsylvania, West Chester, PA 19383, USA}
\altaffiltext{3}{Department of Physics \& Astronomy, University of Leicester, Leicester, LE1 7RH, UK}
\altaffiltext{4}{Large Binocular Telescope Observatory, University
of Arizona, 933 North Cherry Avenue, Tucson, AZ 85721}
\altaffiltext{5}{Department of Astronomy, School of Physics \& Astronomy, 116 Church Street S.E., University of Minnesota, Minneapolis, MN 55455, USA}

\begin{abstract}
We find that the classical nova V723\,Cas (1995) is still an active
X-ray source more than 12 years after outburst and analyze seven
X-ray observations carried out with \swift\ between 2006 January 31
and 2007 December 3. The average count rate is
$0.022\,\pm\,0.01$\,cts\,s$^{-1}$ but the source is variable
within a factor of two of the mean and
does not show any signs of turning off. We present supporting
optical observations which show that between 2001 and 2006 an
underlying hot source was present with steadily increasing
temperature. In order to confirm that the X-ray emission is
from V723\,Cas, we extract a \rosat\ observation taken in
1990 and find that there was no X-ray source at the position
of the nova. The \swift\ XRT spectra resemble those of the
Super Soft X-ray binary Sources (SSS) which is confirmed by
\rxte\ survey data which show no X-ray emission
above 2\,keV between 1996 and 2007. Using blackbody fits we
constrain the effective temperature to between
$T_{\rm eff}=(2.8-3.8)\times10^5$\,K and a bolometric
luminosity $\gtrsim 5\times10^{36}$\,erg\,s$^{-1}$ and
caution that luminosities from blackbodies are generally
overestimated and temperatures underestimated.
We discuss a number of possible explanations for the continuing
X-ray activity, including the intriguing possibility of steady
hydrogen burning due to renewed accretion.
\end{abstract}

\keywords{stars: novae, stars: individual: V723 Cas}

\section{Introduction}

\begin{deluxetable}{ccccc}
\tablecaption{\label{ferats}Journal of Optical observations}
\tablenum{1}
\tablehead{\colhead{Date} & \colhead{UT Time } & \colhead{$t_{\rm exp}$ (s)} &
\colhead{Obs.\tablenotemark{a}} & \colhead{[Fe\,{\sc x}]/[Fe\,{\sc vii}]\tablenotemark{b}}\\
\colhead{} & \colhead{(hh:mm:ss)} & \colhead{} & \colhead{} &
\colhead{}
}
\startdata
2001 Jan 19 & 03:57:21  &  300  &  SO 90'' & 0.02 \\
2003 Sep 13 &  09:39:37  &  600  &  MMT & 0.62 \\
2004 Jun 26 &  10:17:09  &  600  &  SO 90'' & 0.81 \\
2006 Nov 9  &  02:42:59  &  600  &  SO 90'' & 2.33
\enddata
\tablenotetext{a}{Steward Observatory Bok 90 inch telescope and the
Multiple Mirror Telescope.}
\tablenotetext{b}{Ratio of the [Fe\,{\sc x}] (6373\AA) to
[Fe\,{\sc vii}] (6087\AA) lines.}

\end{deluxetable}

\begin{figure*}[!ht]
\resizebox{\hsize}{!}{\includegraphics{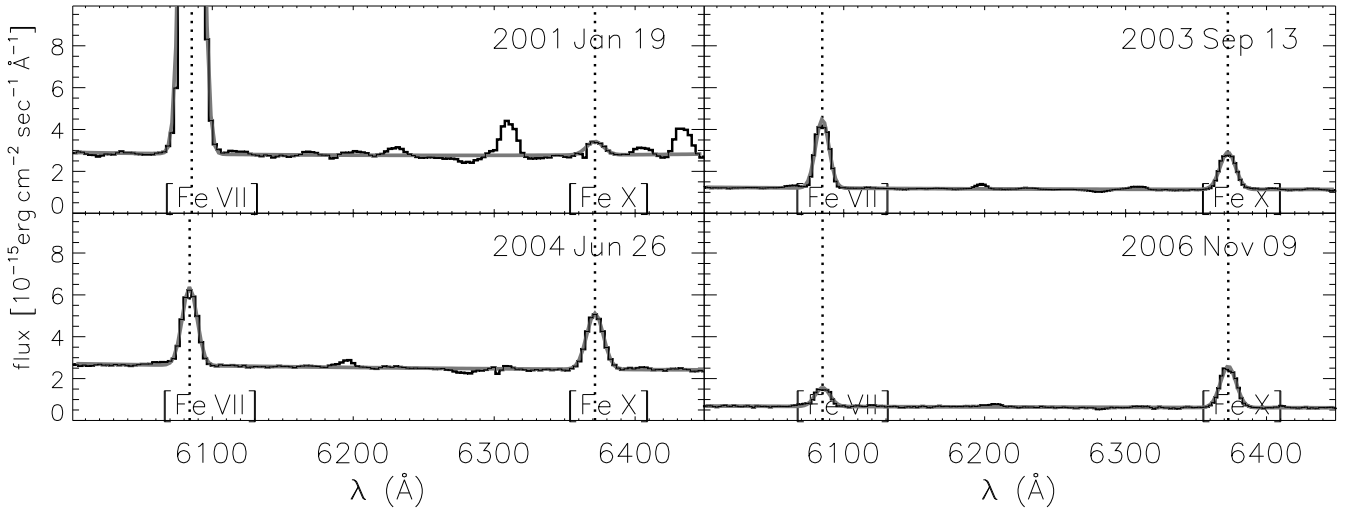}}
\caption{\label{optspec}Optical spectra of V723\,Cas taken between
January 2001 and November 2006, focusing on the $[$Fe\,{\sc x}$]$
and $[$Fe\,{\sc vii}$]$ emission lines at 6376\,\AA\ and 6089\,\AA,
respectively (see Table~\ref{ferats}). The ratio of these lines
is continuously increasing, suggesting that there is a hot source
ionizing increasingly tenuous ejecta.}
\end{figure*}


Classical Novae (CNe) are the observable events caused by
thermonuclear runaways on the surfaces of white dwarfs (WDs),
fueled by material accreted from a companion star. Material
dredged up from below the WD surface is mixed with the accreted
material and violently ejected. The outburst continues until
either nuclear burning has converted all the remaining hydrogen 
on the WD surface to helium or it has been ejected from off
the WD. The bolometric luminosity at the
peak of the explosion is near or exceeds the Eddington luminosity,
$\sim 1\times10^{38}$\,erg\,s$^{-1}$ for a 1-M$_\odot$ WD
\citep[see for example][]{schwarz_lmc}. Initially, the
radiative energy output of the nova peaks in the optical, since
the binary is surrounded by expanding ejecta that are not
transparent to the high-energy radiation produced on the surface
of the WD by nuclear burning. As the ejecta expand and thin
(and the remaining material on the
WD burns hydrostatically in a shell), the photosphere recedes
to inner and hotter layers. The peak of the spectral energy
distribution then shifts to higher energies
\citep{gallstar78}.

 The X-ray bright phase of the evolution has only been observed
sporadically, with the first systematic observations obtained with
\rosat\ \citep[e.g.,][]{kr02}. The X-ray spectrum during this phase
was found to sometimes resemble that of the class of Super Soft
X-ray Binary Sources \citep[SSS;][]{heuvel} such as
\object[Cal 83]{Cal\,83} in the LMC \citep[e.g.,][]{lanz04}.
This phase is now called the SSS phase.
V1974 Cyg was the first nova to be followed
in X-rays from before the SSS phase until decline.
\cite{krautt96} reported a slow rise and fast decline of
the X-ray brightness after the peak had been reached. The first
spectral modeling of V1974 Cyg was carried out by
\cite{krautt96} using blackbody fits, but \cite{balm98}
reanalyzed the spectra using LTE atmosphere models, because
blackbodies did not fit the spectra.
It is further known that luminosities and
temperatures derived from blackbody fits are inconsistent with
realistic physical conditions, with a tendency to overestimate
the luminosity and thus underestimate the temperature
\citep{kahab}.
The LTE models presented by \cite{balm98} were an attempt
to account for complex absorption patterns that change the shape
of the spectrum. Although only 24 spectral bins were available,
an attempt was made to determine a large number of parameters
(temperature, surface gravity, bolometric luminosity, neutral
hydrogen column density, and the abundances of carbon, oxygen,
and neon). Further, LTE and a static WD atmosphere were
assumed. The combination of a small ratio of independent data
points to adjustable parameters and the limitations of current
atmosphere models call for caution when drawing conclusions from
these models. The effects of the
expansion, especially during the early phases of the evolution,
have been investigated with the PHOENIX code by \cite{petz05}.

In addition to the SSS spectrum \cite{krautt96} found a hard
component that was fit with a MEKAL model by \cite{balm98}
and thus was assumed to be an optically thin plasma that is
radiatively cooling. This phenomenon
has been observed in a number of novae \citep[e.g.,][]{swnovae},
the most recent example being RS\,Oph \citep[e.g.,][]{ness_rsoph}.

A systematic search of all \rosat\ observations of novae
including the \rosat\ All Sky Survey and the archives of pointed
\rosat\ observations was carried out by \cite{orio01}. Of a
sample of 108 CNe and recurrent novae (RNe) that exploded
within the last $\sim 100$ years, very few SSS spectra were
found implying that the SSS phase must be short. A recent survey
of all novae observed with \swift\ also resulted in few
detections of CNe in the SSS phase \citep[e.g.,][]{swnovae}. Few
Galactic (e.g., \citealt{orio01,swnovae} and references within)
and extra-galactic \citep{pietsch05,pietsch06} novae have had
the duration of this phase determined. In the extra-galactic
surveys candidates for novae in the SSS phase can be confused
with supernova remnants, foreground neutron stars, or black
hole transients \citep{orio06}.

SSSs are believed to be WDs with
accretion rates high enough to sustain steady nuclear burning
\citep{heuvel,kahab}. Stable shell burning can cause the WD to
grow in mass via accretion with a buildup of hydrogen deficient
material. Thus, the SSSs are strong candidates for
single-degenerate supernovae Ia (SN\,Ia) progenitors
\citep{starrf04}. In general, CNe are not considered to be
SN\,Ia progenitors because the outburst is believed to eject
more material than is accreted.

While theoretical models predict novae to be active for 10s to
100s of years \citep[e.g.,][]{st89,sala06}, all novae observed
so far in X-rays have turned off after much less than a decade
\citep{orio01,swnovae}, implying mass must be lost during the
outburst by some mechanism \citep[e.g.,][]{macdonald85}. The
typical observed outburst duration is of order one year with
the longest observed Galactic nova in outburst being GQ\,Mus
at $\sim$9 years \citep{oegel93,shanley95}.
The outburst duration depends on the nuclear burning rate, the
amount of hydrogen left on the WD after the explosion, and the
rate of any ongoing mass loss. Unfortunately, all these properties
including the WD mass, the most critical driver, are poorly
known.

A different approach was proposed by \cite{greiner03} who found
from the available data that systems with shorter orbital
periods display long durations of supersoft X-ray phases, while
long-period systems show very short or no SSS phase at all.
They speculate that shorter periods may be related to a higher
mass transfer rate (e.g., by increased irradiation) and thus
to the amount of material accreted before the explosion.

In this paper we present \swift\ observations in X-rays of the
slow nova V723\,Cas (\S\ref{back}). In \S\ref{obs} \&
\S\ref{analysis} we describe the observations and our analysis
and address the
possible future X-ray evolution in \S\ref{disc}.\\

\section{V723 Cas background}
\label{back}

\object[V723 Cas]{V723\,Cas} was discovered on 1995 August 24
\citep{1995IAUC.6213....1H}, and reached visual maximum on
1995 December 17 at 7.1\,mag \citep{1996A&A...315..166M}.
Its pre-2001 evolution has been extensively observed from
ultraviolet \citep{1996IAUC.6295....1G} to radio
\citep{2005MNRAS.362..469H} wavelengths. A detailed
description of the optical photometric evolution since 2001
is given by \citet[][ and references therein]{goranskij07}.
The orbital period is $\sim 16.6$ hours with a
sinusoidal-like shape of the V-band variations
\citep{goranskij07}.

We have monitored the optical spectral evolution of V723 Cas
since the outburst. Beginning in 2001, V723\,Cas evolved to
an extreme "coronal" phase with the emergence of the high
ionization line of
$[$\ion{Fe}{10}$]\lambda 6376$\,\AA\ \citep[Fig.~\ref{optspec};
see also ][]{iijima06}.
This iron line was also strong in the late-time spectra of
\object[GQ Mus]{GQ Mus} \citep{1989ApJ...341..968K,oegel93}
which was observed by \rosat\ to have a SSS X-ray spectrum. In
Fig.~\ref{optspec} we show the development of
$[$\ion{Fe}{10}$]$ relative to $[$\ion{Fe}{7}$]
\lambda 6089$\,\AA, between January 2001 and November 2006 in
V723\,Cas (see Table~\ref{tab2}; for details on the observations
we refer to Schwarz et al. in preparation). By 2006 the
$[$\ion{Fe}{10}$]$ line flux exceeded that of $[$\ion{Fe}{7}$]$
indicating that the ejecta were being photoionized by a hot
($>2\times10^5$K) central source, implying ongoing nuclear
burning on the WD \citep[see also][]{greenhouse90}.
The $[$\ion{Fe}{10}$]$ and $[$\ion{Fe}{7}$]$ lines can be
used to monitor the temporal evolution of the photoionizing
X-ray source and constrain the turn-off timescale. The spectral
similarity of V723\,Cas to GQ Mus motivated our initial
\swift\ X-ray observation. Further analysis of the optical
spectra, along with data from other wavelengths, will appear
in an upcoming paper.

\subsection{Extinction and Distance determinations}
\label{ext-dis}

Determining the X-ray luminosity requires knowledge of the
extinction and distance to V723 Cas. The HEASARC $N_{\rm H}$
tool\footnote{http://heasarc.gsfc.nasa.gov/cgi-bin/Tools/w3nh/w3nh.pl}
calculates the total Galactic \ion{H}{1} column density for any direction
using the Leiden/Argentine/Bonn \citep[LAB; ][]{Kal05} and \citet{dickey90}
Galactic \ion{H}{1} surveys. For a cone of radius $0.5\arcdeg$ centered
on the J(2000) coordinates of V723 Cas, the LAB and \citet{dickey90} maps
give average $N_{\rm H}$ values of $2.2\times10^{21}$\,cm$^{-2}$ and
$2.5\times10^{21}$\,cm$^{-2}$, respectively. 
Since there is a possibility of circumstellar hydrogen from
either the ejecta or more recent mass loss, direct
measurements of extinction are warranted.
We use estimates of E(B-V) from optical and UV data to derive $N_{\rm H}$.
\citet{chochol97} found E(B-V) of $\simeq 0.59$ and $\simeq 0.54$ from
the observed versus intrinsic (B-V) values at maximum and 2 magnitudes
after maximum. \citet{1996A&A...315..166M} estimated E(B-V)
$\simeq 0.45$ from a fit to the early interstellar \ion{Na}{1} D
absorption lines, while the \citet{schlegel98} extinction maps give
an E(B-V) of $\simeq 0.4$ for the V723 Cas location. The
2175\AA\ feature in the UV gives another direct measurement of E(B-V).
Early IUE observations, while the source was still optically thick,
indicated E(B-V) = 0.6 \citep{1996IAUC.6295....1G} while a recent \galex\
observation implies a slightly lower value of 0.5 \citep{swaas07}.
All methods are consistent with E(B-V)$=0.5\pm0.1$. Converting
to $N_{\rm H}$ using the relation
$<N_{\rm H}$/E(B-V)$>=6\,\pm\,2\times10^{21}$\,cm$^{-2}$
\citep{dickey90,bohlin78}
constrains $N_{\rm H}$ to values between $1.6\times10^{21}$\,cm$^{-2}$ and
$4.8\times10^{21}$\,cm$^{-2}$ which is consistent with the Galactic \ion{H}{1}
column density maps.

There are three distance estimates in the literature.
\citet{2005MNRAS.362..469H} used model fits to the radio light curve to
obtain $d = 2.4\,\pm\,0.4$\,kpc. \citet{iijima98} estimated a distance
of 2.95\,kpc from the strength of the interstellar \ion{Na}{1} D
absorption lines in the early spectra while \citet{iijima06} applied a
maximum magnitude versus rate of decline (MMRD) relationship to the
smoothed light curve for a distance of 2.8 kpc. Since the early
light curve of V723\,Cas was extremely irregular, the MMRD method
is problematic 
\citep[see also ][]{schwarz07}. Another approach is to assume
that novae with similar characteristics, e.g., light curve evolution,
ejection velocities, etc., have the same absolute magnitudes at maximum.
For V723 Cas we use the similar slow nova, \object[HR Del]{HR\,Del},
which has an accurate distance estimate from expansion parallax measurements
of its ejecta \citep{HO03}. The distance from this assumption,
allowing for the uncertainty in the reddening given above, is
$2.7^{+0.4}_{-0.3}$\,kpc.
All four estimates are consistent and thus we adopt $d=2.7$\,kpc
for the rest of the paper.

\section{Observations}
\label{obs}

\begin{figure}[!ht] 
\resizebox{\hsize}{!}{\includegraphics{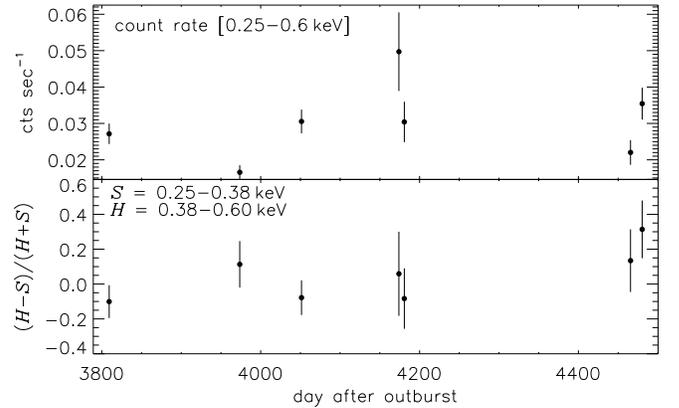}}
\caption{\label{hr_time}Count rates (top) and hardness
ratios (bottom) as a function of time (day after outburst).
}
\end{figure}

\begin{deluxetable}{llclr}
\tablecaption{\swift\ XRT Observations}
\tablenum{2}
\label{tab2}
\tablehead{
\colhead{Date} & \colhead{ObsID} &
\colhead{$t_{\rm exp}$ (s)} & \colhead{$10^{-3}$\,cts\,s$^{-1}$}
& \colhead{hardness}
}
\startdata
2006 Jan 31.3 & 00030361001    & 6840 & $26.3\,\pm\,2.2$ & $-0.10\,\pm\,0.09$\\
2006 July$^a$ & 0003036100$^a$ & 8650 & $14.0\,\pm\,1.0$ & $0.11\,\pm\,0.13$  \\
2006 Sep 30.4 & 00030361008    & 5750 & $30.1\,\pm\,2.5$ & $-0.08\,\pm\,0.10$ \\
2007 Jan 31.0 & 00030361010    & 900  & $49.2\,\pm\,8.7$ & $0.06\,\pm\,0.24$  \\
2007 Feb 06.8 & 00030361011    & 1970 & $29.9\,\pm\,4.5$ & $-0.08\,\pm\,0.17$ \\
2007 Nov 18.6 & 00030361012    & 3800 & $21.7\,\pm\,2.7$ & $0.13\,\pm\,0.18$  \\
2007 Dec 03.3 & 00030361013    & 3600 & $35.0\,\pm\,3.4$ & $0.31\,\pm\,0.17$
\enddata
\tablenotetext{a}{Four observations between July 9.1 and July 14.7,\\
ObsIDs 0003036100[3,5,6,7]}
\end{deluxetable}

\begin{figure}[!ht]
\resizebox{\hsize}{!}{\includegraphics{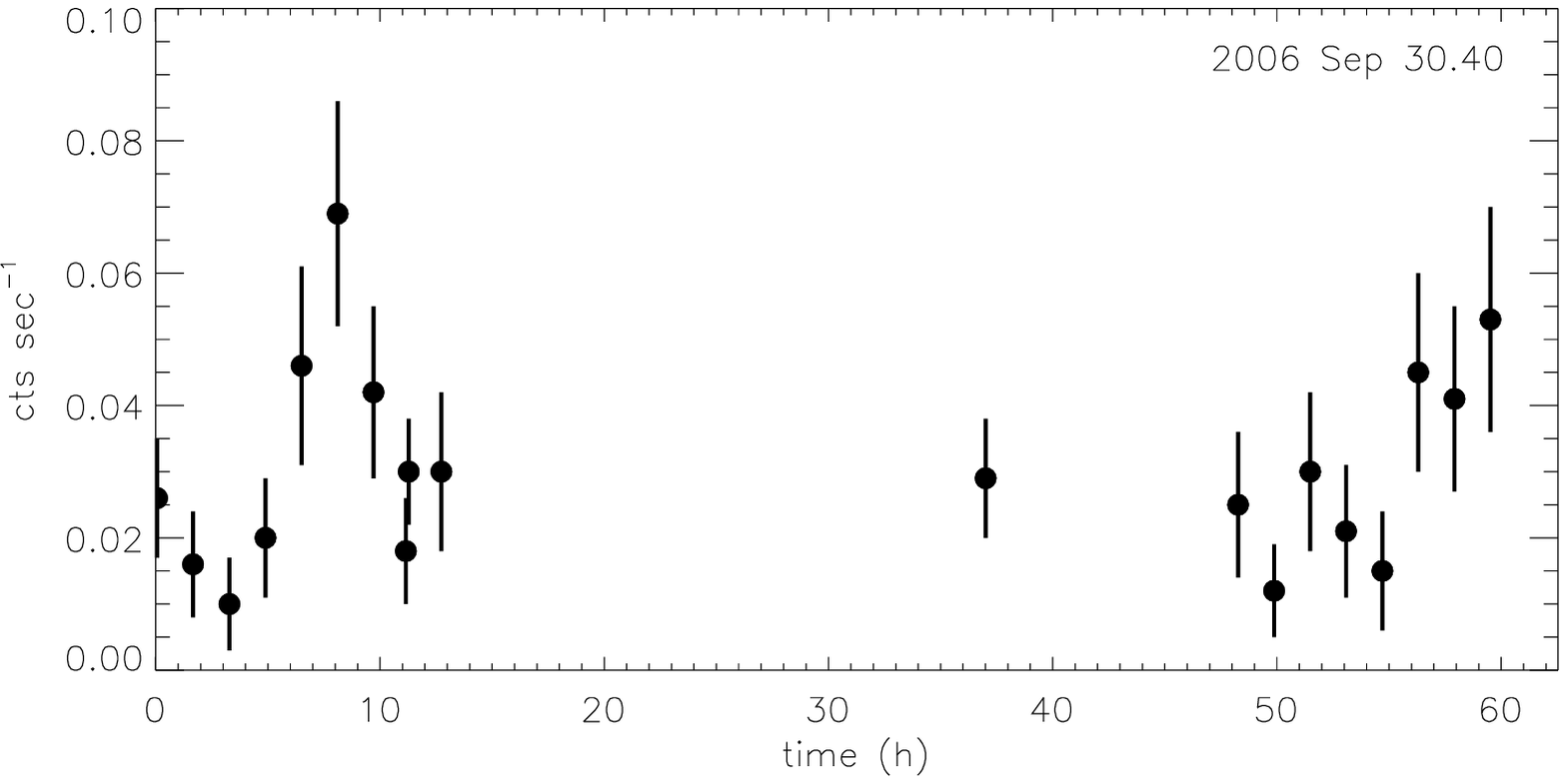}}

\resizebox{\hsize}{!}{\includegraphics{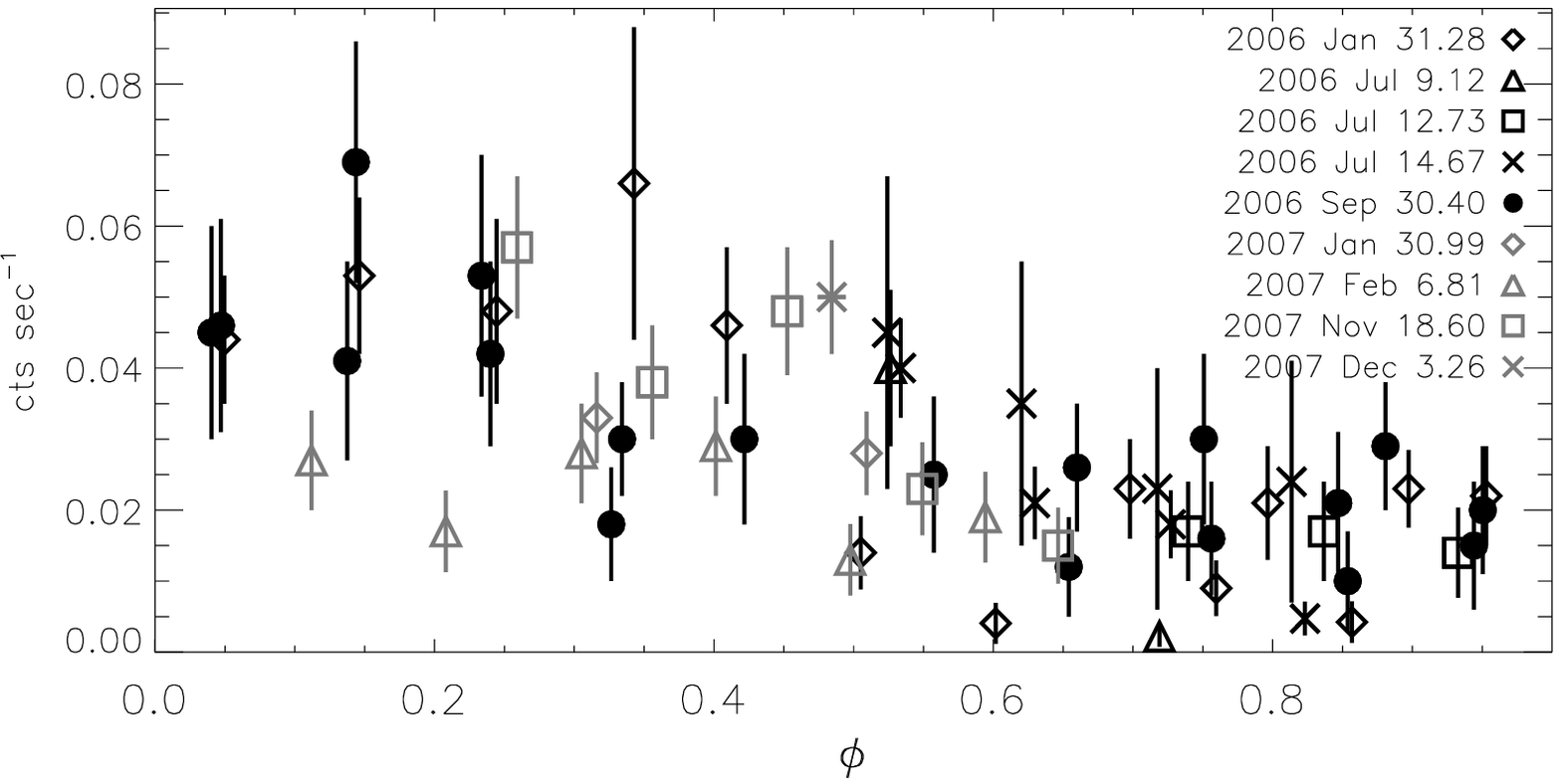}}
\caption{\label{lc}{\bf Top}: Light curve for ObsID
00030361008, extracted from 0.25-10\,keV. The
observation consists of 19 observation
intervals (OBIs) and considerable variability
can be identified. {\bf Bottom}: Average
count rates extracted from all OBIs and plotted versus
orbital phase \citep[$\phi=0$ marks optical
minimum; ephemeris taken from][]{goranskij07}.
}
\end{figure}

The first observation with the \swift\ X-Ray Telescope
\citep[XRT;][]{xrt} of V723\,Cas was taken on 2006
January 31.27 and yielded a clear detection with no
counts above 0.6\,keV and a peak at $\sim 0.35$\,keV,
thus typical of a SSS spectrum
\citep{2006IAUC.8676....2N,swnovae}. We have
obtained six more \swift\ observations on the dates listed
in Table~\ref{tab2}, and the mean count rate for all
observations is $0.022\,\pm\,0.01$\,cts\,s$^{-1}$.
In this table we also list the ObsIDs, exposure times,
measured count rates and spectral hardness ratios,
$HR=(H-S)/(H+S)$ with $S$ and $H$ denoting the
count rates in the energy ranges 0.25-0.38\,keV and
0.38-0.60\,keV, respectively. The
extraction of count rates and spectra is described in
\cite{swnovae}. Briefly, we use a circular extraction region
with radius 10 pixels for the source and an annular
extraction region with inner radius 10 pixels and
outer radius 100 pixels for the background. The
extracted count rates are independent of the size
of the extraction region. In Fig.~\ref{hr_time} we
show the evolution of count rate and hardness
ratio with time. While the count rate varies by over
a factor of two from the mean, no changes of
the hardness ratio can be established at a
significant level (see also Table~\ref{tab2}).

Due to the orbit of the space craft, only short continuous
observations (snapshots) can be taken. Each observation
consists of between 1 and 20 short snapshots separated by
hours to days. We extracted the count rates for each
snapshot individually and show the evolution of count
rate for ObsID 00030361008 in the top panel of Fig.~\ref{lc}.
During the
first part of the observation a flare with an amplitude of
a factor three above the pre-flare and post-flare rates can
be seen. The other observations also show considerable
variability. In the bottom panel of Fig.~\ref{lc} we show
all observed count rates as a function of orbital phase, which was
computed based on the ephemerides given by \cite{goranskij07}.
While the count rates between phase $\phi=0$ and 0.5 appear to be
slightly higher, there is no systematic trend with phase
in this dataset.

 Next we extracted spectra for each observation and found
all to be typical SSS spectra.
Since the hardness ratio doesn't change at a significant level,
no significant difference
between model parameters related to the spectral shape (e.g.,
temperature) can be expected, and we combined all observations
into a single spectrum (31.4\,ksec) for the spectral analysis
(see Fig.~\ref{bb}).

V723\,Cas was also detected in the \xmm\ Slew Survey Full
Source Catalog \citep[v1.1,][]{xmmsurvey} in an
observation taken on 2007 February 1 (22:33:51 start time;
ObsID 9130900003) with a count rate of 1.2\,cts\,s$^{-1}$.
We converted this count rate to an equivalent \swift\ XRT
rate using the HEASARC tool {\tt PIMMS}, assuming a blackbody
spectrum with k$T=30$\,eV, and the converted value is consistent
with our contemporaneous observations. There is no
detection in the hard energy band, defined as 2--12\,keV.

As a confirmation that the \swift\ X-ray source is associated
with the 1995 outburst of V723 Cas, we inspected the \rosat\
archive for any observations prior to August 1995 near the
coordinates of V723\,Cas. We found one match, an observation
on 1990 July 30 (ObsID 930702) from which we determine
an upper limit of $4.4\times10^{-4}$\,cts\,s$^{-1}$ at the
coordinates of V723 Cas. Converting the mean \swift\ count
rate to \rosat\ counts with {\tt PIMMS}, results in an expected
\rosat\ count rate $\approx 1000$ times brighter than the
measured upper limit. Thus we conclude that
there was no X-ray source at this position prior to the 1995
outburst and that the current \swift\ observations are
associated with the 1995 outburst of V723 Cas.

V723 Cas was included in the survey of \rxte\ All-Sky Monitor
Long-Term Observed Sources. Between 1996 January 4 and
2007 September 20, 75691 observations were taken in the energy
band between 2 and 12 keV with exposure times between 11 and 92
seconds. Out of this sample there are
30 detections at a significance level $>3\sigma$ and five at
$>4\sigma$, with no systematic trend in time.
Meanwhile, there are 2793 observations where the
background-subtracted count rate is below zero at the
$>3\sigma$ significance level, and five events have negative
count rates at a 4-$\sigma$ significance level. We conclude
that none of these observations deliver convincing detections
of the source in hard X-rays. This assertion is consistent
with the non-detection of V723\,Cas in the 2--12-keV band
of the \xmm\ observation and with our soft \swift\ spectra
yielding no signal above 0.6\,keV (Fig.~\ref{bb}).

\section{Analysis}
\label{analysis}

Given the relatively modest spectral resolution of the XRT,
we utilized
blackbody fits, $B_\lambda(T_{\rm eff})$, to determine the
range of temperature consistent with the observed spectrum.
We took the temperatures, $T_{\rm eff}$, and the emitting
radii, $R_{\rm EM}$ (derived from the normalization assuming
a distance of $d=2.7$\,kpc), to determine bolometric luminosities,
$L_{\rm bol}$. We corrected for interstellar
absorption by fitting a parameterized bound-free absorption model
implemented in \mbox{PINTofAle} \citep{pintofale}. The total
column density from free-bound absorption by neutral elements
in the line of sight was computed from the hydrogen
column density, $N_{\rm H}$, assuming solar abundances.
In an iterative process we determined the combination of the
three parameters $T_{\rm eff}$, $R_{\rm EM}$, and $N_{\rm H}$
that yielded the best agreement with the measured spectrum,
using a maximum likelihood estimator.
We determined uncertainties of the parameters
by application of the likelihood ratio test, i.e., we consider
two models to be different with probabilities 68.3\%,
84.3\%, 95.45\%, and 99\% if the likelihood,
increases by 3.53, 6.25, 8.02, and 11.3, respectively, from the
lowest of all likelihood values (for three free parameters).
We carried out rigorous optimization of all three
parameters, but found an unrealistically high value of
luminosity and a value of $N_{\rm H}$ that is inconsistent
with the \galex\ observations (\S\ref{ext-dis}). We
therefore explore the full uncertainty range and constrain 
the range of solutions within these uncertainties plus
physical arguments.

\begin{figure*}[!ht]
\resizebox{\hsize}{!}{\includegraphics{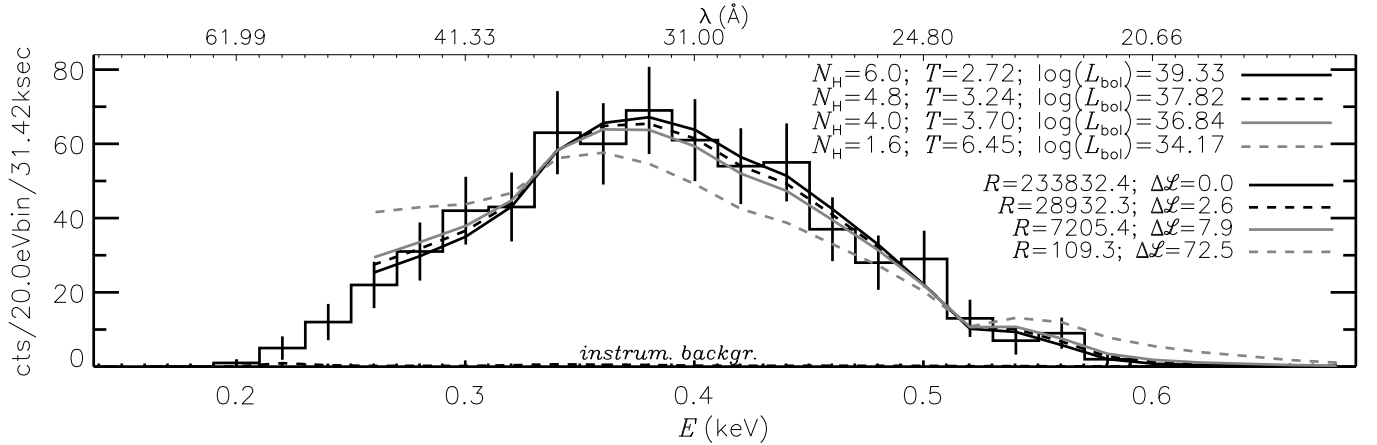}}
\caption{\label{bb} Blackbody fits to the combined XRT spectrum
(rebinned by a factor of two). Spectral bins below 0.25\,keV
were excluded from the fit because of uncertain XRT calibration.
We tested four different values of the hydrogen column density,
$N_{\rm H}$ (given in $10^{21}$\,cm$^{-2}$), and optimized
the effective temperature (T given in $10^5$\,K) and
bolometric luminosity ($\log L_{\rm bol}$ in erg\,s$^{-1}$).
Emitting radii, R, are given in km. The values of
$\Delta\Ell$ in the legend indicate the change in
likelihood from the fit with the lowest likelihood
value (set to 0). Note that the fit associated with the lowest
value of $\Ell$ is in contradiction to the independent measurements of
$N_{\rm H}$ and is physically unrealistic.}
\end{figure*}

In Fig.~\ref{bb} we compare four fits with fixed
$N_{\rm H}=(6.0,\,4.8,\,4.0,\,1.6)\times10^{21}$\,cm$^{-2}$
but fitted
$T_{\rm eff}$ and $R_{\rm EM}$ and find the values
listed in the legend. In the bottom right we show the likelihood values
compared to the fit with the lowest likelihood.
It can be seen that better fits are achieved with values of
$N_{\rm H}$ greater than the upper bound found from the \galex\
observations (see \S\ref{ext-dis}). However, the
likelihood for the fit with $N_{\rm H}=4.8\times10^{21}$\,cm$^{-2}$
increases by less than 3.53 in value and is thus consistent within 68.3\,per
cent propability. In addition to statistical arguments from
the spectral fitting, physical arguments must be considered. For
example, the fit with $N_{\rm H}=1.6\times10^{21}$\,cm$^{-2}$
gives a radius that is too small for a WD, while the fit with
$N_{\rm H}=6.0\times10^{21}$\,cm$^{-2}$ is too luminous.
The super-Eddington
luminosity phase in novae is extremely short, lasting at best
days and certainly not over 12 years \citep{schwarz_lmc}.

\begin{figure*}[!ht]
\resizebox{\hsize}{!}{\includegraphics{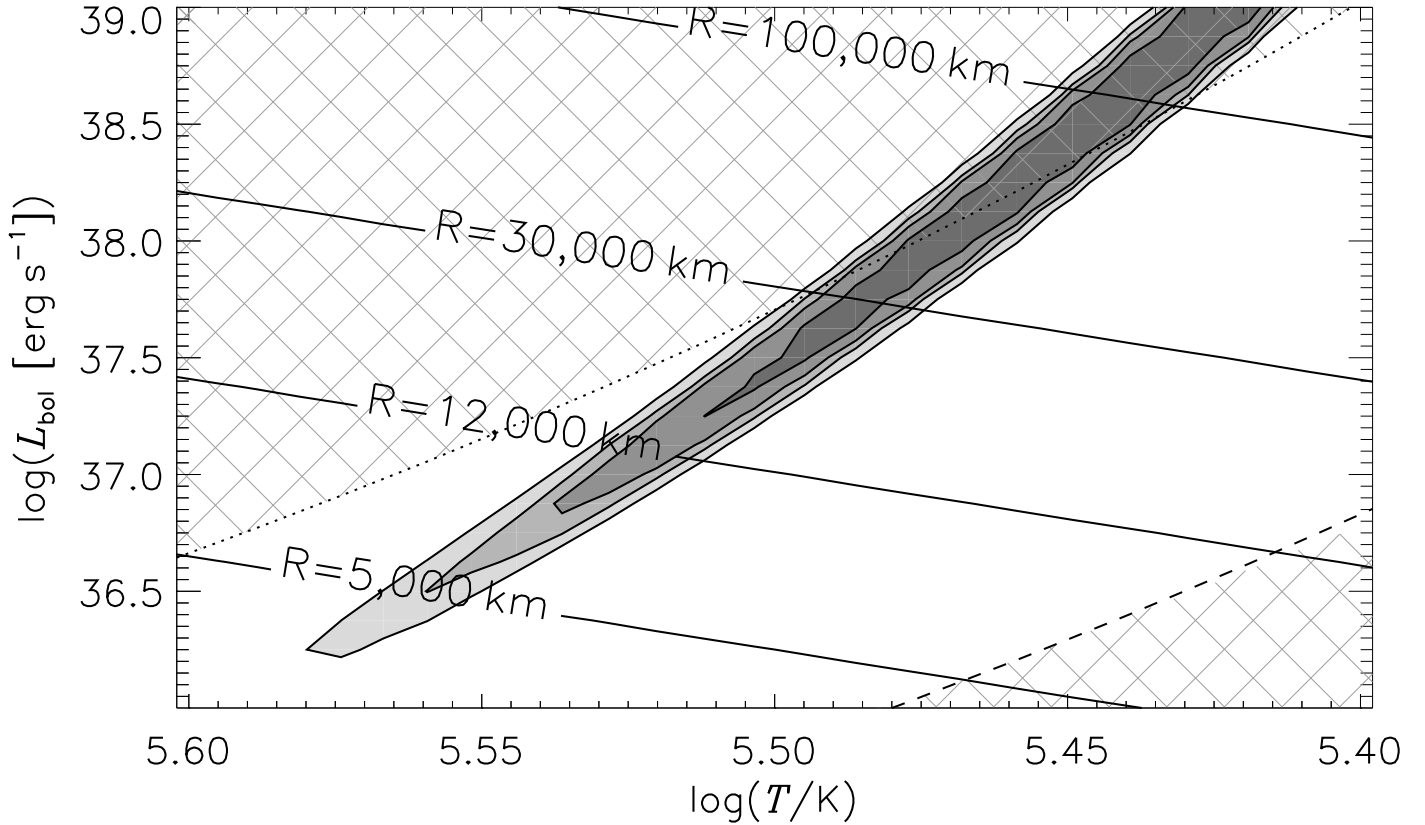}}
\caption{\label{contour} Likelihood contours for the
temperature and luminosity obtained from blackbody fits
to the combined XRT spectrum of V723\,Cas. From dark to
light shades the contours indicate the levels of 68.3\%,
84.3\%, 95.45\%, and 99\% likelihood. We draw lines of
constant emitting radii $R_{\rm EM}$. The highest and
lowest values of $N_{\rm H}$ determined from
independent measurements are indicated with dotted and
dashed lines, respectively (values from \S\ref{ext-dis}),
and we cross out regions that are inconsistent with these
measurements. All fits that are consistent with the data
within 99\% confidence and with the independent
$N_{\rm H}$ measurements yield luminosities
below the Eddington limit.
}
\end{figure*}

In Fig.~\ref{contour} we show the results from computing a grid
of models as a contour plot with the confidence intervals from
the likelihood ratio test shown in shades of gray. Overplotted
are the contours of the upper and lower $N_{\rm H}$ values
derived in \S\ref{ext-dis} and contours of emitting radii.
The $R_{\rm EM}$ values can be interpreted as
a proxy of the WD mass. A static 1-M$_\odot$ WD has a radius of
$\sim 6,000$\,km \citep[e.g., Sirius B; ][]{Bar05}. WDs with
a lower mass will have a larger radius, or the
radius can be larger as a result of surface hydrogen burning.

A 1.4-M$_\odot$ WD has a radius of $\sim 2000$\,km
(computed from eq.~3 in \citealt{truranlivio86}), and no radii
below $R_{\rm EM} = 2,000$\,km are physically realistic.
We also have to exclude luminosities above the Eddington limit,
which is $\sim 2\times 10^{38}$\,erg\,s$^{-1}$ for
a 1-M$_\odot$ WD. The most likely region in
the luminosity-$T_{\rm eff}$ diagram shown in Fig.~\ref{contour}
is therefore the 3-$\sigma$ contour area,
 bounded by the
$N_{\rm H} = 4.8\times10^{21}$\,cm$^{-2}$ contour on the left
and top. From Fig.~\ref{contour} it is apparent that this
boundary also excludes the physically unrealistic luminosities
above log($L_{\rm bol}$) = 38.3. Meanwhile, the lower bound of
$1.6\times10^{21}$\,cm$^{-2}$ is less likely, according to our
fits than the higher value of $N_{\rm H} = 4.8\times10^{21}$\,cm$^{-2}$.
With these boundaries, we find a range in $T_{\rm eff}
\sim (2.8-3.8)\times10^5$\,K and $L_{\rm bol}$ greater than
$5\times10^{36}$\,erg\,s$^{-1}$, but less than
$2\times10^{38}$\,erg\,s$^{-1}$. Within these uncertainties
and those in the distance, the derived V band magnitude from the
range of blackbody fits is consistent with the current
V band photometry.

In an attempt to test the viability of our blackbody
fits, we have applied the same method to a \chandra\
LETGS spectrum of the canonical SSS Cal\,83, which
is sufficiently well exposed and has enough spectral
resolution for an atmosphere analysis \citep{lanz04}.
We found a significantly
higher luminosity and lower temperature than the atmosphere
results. This confirms that blackbody fits
systematically overestimate the luminosity and underestimate
the temperature \citep[see also, e.g.,][]{greiner91}
suggesting that the actual luminosity is lower and the
temperature is higher.

\section{Discussion}
\label{disc}

The long life time of V723\,Cas is a significant departure from
the typical behavior of Galactic CNe. It requires that a
reservoir of hydrogen-rich material be maintained over
timescales much longer than usual. For normal CNe this can be
done with either a low-mass WD leading to a lower burning rate, a
large amount of hydrogen left on the WD after the outburst, or
inefficient mass loss mechanisms (i.e. winds, common envelope
frictional drag, etc.) after outburst. With a conversion
efficiency of $6\times 10^{18}$\,erg\,g$^{-1}$ and a constant
luminosity of $10^{37}$\,erg\,s$^{-1}$, only $3\times
10^{-7}$\,M$_\odot$ of hydrogen is required to power the nova
over a 12 year period, assuming no mass loss. Lower luminosities
could further extend the duration of nuclear burning. Regardless
of the combination of factors involved, if V723\,Cas is only
burning residual accreted hydrogen, it will eventually turn off.

 Other systems with long outburst lifetimes such as
\object{AG Peg} and \object{RR Tel} \citep{ness_rrtel} are all
Symbiotic Novae which are different from CNe because they have
longer orbital
periods (hundreds of days) and giant secondaries with large
wind mass loss rates. The WDs of these long-lived systems are
probably of lower mass than those in CNe (e.g, RR Tel;
\citealt{rrtel_rosat})
which allows a larger envelope of hydrogen-rich material prior
to outburst. The turnoff of the hydrogen
burning shell occurs on a nuclear burning timescale
rather than the faster hydrogen envelope depletion timescale
which includes mass loss from the ejection of previously
accreted material and wind mass loss \citep{macdonald85}.
The temperature of the
WD in RR Tel is lower than the one we find for V723\,Cas
or for the canonical SSS Cal\,83 (\citealt{rrtel_rosat} found
$T_{\rm eff}=142,000$\,K), again showing that Symbiotics
are different from CNe and SSSs.
V723\,Cas is not a Symbiotic Nova, since the orbital period
is much shorter, and the companion is not a giant, which,
at a distance of 2.7\,kpc and E(B-V)=0.5 (\S\ref{ext-dis})
would lead to a higher observed magnitude than $V\sim15$.
In addition, a recent optical spectrum of V723\,Cas taken
during the optical minimum revealed no signature of a
late-type, giant companion (Schwarz et al. in preparation).
In order for the V723\,Cas secondary to fill its Roche
lobe, it must be a subgiant. The majority of CNe have
shorter orbital periods.

There is, therefore, the interesting possibility that
accretion has been re-established in the V723\,Cas system and
is supplying the WD with additional fuel.
No signs of V723\,Cas
to be an intermediate polar (IP) can be found. IPs
typically have strong hard X-ray components as in AM\,Her's
and at least two periods in their optical light curves,
the orbital and the spin period of the WD. In fig.~5 of
\cite{swnovae} the X-ray spectra of V723\,Cas and the
IP V4743\,Sgr are compared. Although V4743\,Sgr is
further away, the X-ray count rate at high energies is
much higher, while there are no counts at high energies
that can be attributed to V723\,Cas. Further, a strong
period of 22\,min was found in the X-ray light curve
of V4743\,Sgr \citep{v4743}. The only optical period
for V723\,Cas seems to be the orbital period
of $\sim 16.6$ hours \citep{goranskij00}. Also,
\cite{goranskij07} report of no other period that
would indicate an IP.

\section{Summary and Conclusions}
\label{conc}

Although other CNe have been observed to be active for several
years, V723\,Cas is now confirmed to be the longest CN observed
in outburst and as of December 2007 shows no indication it is
turning off. While the observed count rate was variable, the
spectral hardness remained unchanged. 
From blackbody fits we find a temperature range from 2.8$\times$10$^5$ to
3.8$\times$10$^5$\,K and the luminosity greater than
5$\times$$10^{36}$\,erg\,s$^{-1}$. We caution that these numbers
have limited physical meaning, as it is well known that blackbody
fits lead to overestimated luminosities and thus underestimated
temperatures \citep{kahab}. Nevertheless, the high temperatures
(especially if the numbers are underestimated) indicate that nuclear
burning is still continuing. V723\,Cas may be burning
the remaining hydrogen left on the WD surface after
the initial explosion. If so, the X-ray emission will decline
once the fuel is exhausted. It is unknown when this event will
happen but we plan on continuing our \swift\
monitoring program to detect a possible X-ray turn off and
to further investigate the changes in the X-ray count rate.

 There is also the intriguing possibility that the nova is
now fed via renewed accretion similar to that observed
in the prototype SSS, Cal\,83. If so, V723 Cas may have
evolved into a permanent SSS. Evidence that points to this
possibility is
the exceptional length of time in outburst and the long
orbital period similar to that in Cal\,83 and Cal\,87.
Note also that \cite{greiner03} found that long-period
systems show shorter SSS phases than short-period systems.
V723\,Cas does not fit into this picture, suggesting that
this is a different situation.
Establishing the WD mass and determining whether accretion has
re-established will assist in understanding the fate of
V723 Cas. Schwarz et al.
(in preparation) are presently obtaining optical spectra during
the different phases of the orbit to determine whether the
V723\,Cas system is dominated by an illuminated subgiant
secondary, a re-established accretion disk, or some mixture
of both. Any orbital variations in the lines may also be used
to provide mass estimates of the system components.

\acknowledgments

J.-U.N. gratefully acknowledges support provided by NASA
through Chandra Postdoctoral Fellowship grant PF5-60039
awarded by the Chandra X-ray Center, which is operated by
the Smithsonian Astrophysical Observatory for NASA under
contract NAS8-03060.
S.S. received partial support from NSF and NASA grants to ASU.
J.P.O., K.L.P, and A.P.B. acknowledge the support of PPARC.
We thank N. Gehrels for granting the observations.

\small
\bibliographystyle{apj}
\bibliography{mybibfile}

\begin{thebibliography}{51}
\expandafter\ifx\csname natexlab\endcsname\relax\def\natexlab#1{#1}\fi

\bibitem[{{Balman} {et~al.}(1998){Balman}, {Krautter}, \& {\"Ogelman}}]{balm98}
{Balman}, S., {Krautter}, J., \& {\"Ogelman}, H. 1998, ApJ, 499, 395

\bibitem[{{Barstow} {et~al.}(2005){Barstow}, {Bond}, {Holberg}, {Burleigh},
  {Hubeny}, \& {Koester}}]{Bar05}
{Barstow}, M.~A., {Bond}, H.~E., {Holberg}, J.~B., {Burleigh}, M.~R., {Hubeny},
  I., \& {Koester}, D. 2005, MNRAS, 362, 1134

\bibitem[{{Bohlin} {et~al.}(1978){Bohlin}, {Savage}, \& {Drake}}]{bohlin78}
{Bohlin}, R.~C., {Savage}, B.~D., \& {Drake}, J.~F. 1978, ApJ, 224, 132

\bibitem[{{Burrows} {et~al.}(2005){Burrows}, {Hill}, {Nousek}, {Kennea},
  {Wells}, {Osborne}, {Abbey}, {Beardmore}, {Mukerjee}, {Short}, {Chincarini},
  {Campana}, {Citterio}, {Moretti}, {Pagani}, {Tagliaferri}, {Giommi},
  {Capalbi}, {Tamburelli}, {Angelini}, {Cusumano}, {Br{\"a}uninger}, {Burkert},
  \& {Hartner}}]{xrt}
{Burrows}, D.~N., {Hill}, J.~E., {Nousek}, J.~A., {Kennea}, J.~A., {Wells}, A.,
  {Osborne}, J.~P., {Abbey}, A.~F., {Beardmore}, A., {Mukerjee}, K., {Short},
  A.~D.~T., {Chincarini}, G., {Campana}, S., {Citterio}, O., {Moretti}, A.,
  {Pagani}, C., {Tagliaferri}, G., {Giommi}, P., {Capalbi}, M., {Tamburelli},
  F., {Angelini}, L., {Cusumano}, G., {Br{\"a}uninger}, H.~W., {Burkert}, W.,
  \& {Hartner}, G.~D. 2005, Space Science Reviews, 120, 165

\bibitem[{{Chochol} \& {Pribulla}(1997)}]{chochol97}
{Chochol}, D., \& {Pribulla}, T. 1997, Contrib. Astron. Obs. Skalnate Pleso,
  27, 53

\bibitem[{{Dickey} \& {Lockman}(1990)}]{dickey90}
{Dickey}, J.~M., \& {Lockman}, F.~J. 1990, ARAA, 28, 215

\bibitem[{{Gallagher} \& {Starrfield}(1978)}]{gallstar78}
{Gallagher}, J.~S., \& {Starrfield}, S. 1978, ARAA, 16, 171

\bibitem[{{G{\"a}nsicke} {et~al.}(1998){G{\"a}nsicke}, {van Teeseling},
  {Beuermann}, \& {de Martino}}]{gaensicke98}
{G{\"a}nsicke}, B.~T., {van Teeseling}, A., {Beuermann}, K., \& {de Martino},
  D. 1998, A\&A, 333, 163

\bibitem[{{Gonzalez-Riestra} {et~al.}(1996){Gonzalez-Riestra}, {Shore},
  {Starrfield}, \& {Krautter}}]{1996IAUC.6295....1G}
{Gonzalez-Riestra}, R., {Shore}, S.~N., {Starrfield}, S., \& {Krautter}, J.
  1996, IAU Circ, 6295, 1

\bibitem[{{Goranskij} {et~al.}(2007){Goranskij}, {Katysheva}, {Kusakin},
  {Metlova}, {Pogrosheva}, {Shugarov}, {Barsukova}, {Fabrika}, {Borisov},
  {Burenkov}, {Pramsky}, {Karitskaya}, \& {Retter}}]{goranskij07}
{Goranskij}, V.~P., {Katysheva}, N.~A., {Kusakin}, A.~V., {Metlova}, N.~V.,
  {Pogrosheva}, T.~M., {Shugarov}, S.~Y., {Barsukova}, E.~A., {Fabrika}, S.~N.,
  {Borisov}, N.~V., {Burenkov}, A.~N., {Pramsky}, A.~G., {Karitskaya}, E.~A.,
  \& {Retter}, A. 2007, Astrophysical Bulletin, 62, 125

\bibitem[{{Goranskij} {et~al.}(2000){Goranskij}, {Shugarov}, {Katysheva},
  {Shemmer}, {Retter}, {Chochol}, \& {Pribulla}}]{goranskij00}
{Goranskij}, V.~P., {Shugarov}, S.~Y., {Katysheva}, N.~A., {Shemmer}, O.,
  {Retter}, A., {Chochol}, D., \& {Pribulla}, T. 2000, Informational Bulletin
  on Variable Stars, 4852, 1

\bibitem[{{Greenhouse} {et~al.}(1990){Greenhouse}, {Grasdalen}, {Woodward},
  {Benson}, {Gehrz}, {Rosenthal}, \& {Skrutskie}}]{greenhouse90}
{Greenhouse}, M.~A., {Grasdalen}, G.~L., {Woodward}, C.~E., {Benson}, J.,
  {Gehrz}, R.~D., {Rosenthal}, E., \& {Skrutskie}, M.~F. 1990, ApJ, 352, 307

\bibitem[{{Greiner} {et~al.}(1991){Greiner}, {Hasinger}, \&
  {Kahabka}}]{greiner91}
{Greiner}, J., {Hasinger}, G., \& {Kahabka}, P. 1991, A\&A, 246, L17

\bibitem[{{Greiner} {et~al.}(2003){Greiner}, {Orio}, \& {Schartel}}]{greiner03}
{Greiner}, J., {Orio}, M., \& {Schartel}, N. 2003, A\&A, 405, 703

\bibitem[{{Harman} \& {O'Brien}(2003)}]{HO03}
{Harman}, D.~J., \& {O'Brien}, T.~J. 2003, MNRAS, 344, 1219

\bibitem[{{Heywood} {et~al.}(2005){Heywood}, {O'Brien}, {Eyres}, {Bode}, \&
  {Davis}}]{2005MNRAS.362..469H}
{Heywood}, I., {O'Brien}, T.~J., {Eyres}, S.~P.~S., {Bode}, M.~F., \& {Davis},
  R.~J. 2005, MNRAS, 362, 469

\bibitem[{{Hirosawa} {et~al.}(1995){Hirosawa}, {Yamamoto}, {Nakano}, {Kojima},
  {Iida}, {Sugie}, {Takahashi}, \& {Williams}}]{1995IAUC.6213....1H}
{Hirosawa}, K., {Yamamoto}, M., {Nakano}, S., {Kojima}, T., {Iida}, M.,
  {Sugie}, A., {Takahashi}, S., \& {Williams}, G.~V. 1995, IAU Circ, 6213, 1

\bibitem[{{Iijima}(2006)}]{iijima06}
{Iijima}, T. 2006, A\&A, 451, 563

\bibitem[{{Iijima} {et~al.}(1998){Iijima}, {Rosino}, \& {della
  Valle}}]{iijima98}
{Iijima}, T., {Rosino}, L., \& {della Valle}, M. 1998, A\&A, 338, 1006

\bibitem[{{Jordan} {et~al.}(1994){Jordan}, {Murset}, \& {Werner}}]{rrtel_rosat}
{Jordan}, S., {Murset}, U., \& {Werner}, K. 1994, A\&A, 283, 475

\bibitem[{{Kahabka} \& {van den Heuvel}(1997)}]{kahab}
{Kahabka}, P., \& {van den Heuvel}, E.~P.~J. 1997, ARA\&A, 35, 69

\bibitem[{{Kalberla} {et~al.}(2005){Kalberla}, {Burton}, {Hartmann}, {Arnal},
  {Bajaja}, {Morras}, \& {P{\"o}ppel}}]{Kal05}
{Kalberla}, P.~M.~W., {Burton}, W.~B., {Hartmann}, D., {Arnal}, E.~M.,
  {Bajaja}, E., {Morras}, R., \& {P{\"o}ppel}, W.~G.~L. 2005, A\&A, 440, 775

\bibitem[{{Kashyap} \& {Drake}(2000)}]{pintofale}
{Kashyap}, V., \& {Drake}, J.~J. 2000, Bulletin of the Astronomical Society of
  India, 28, 475

\bibitem[{{Krautter}(2002)}]{kr02}
{Krautter}, J. 2002, in {AIP Conf. Proc.: Classical Nova Explosions}, ed.
  M.~Hernanz \& J.~Jos\'{e}, Vol. 637 (American Institute of Physics), 345

\bibitem[{{Krautter} {et~al.}(1996){Krautter}, {\"Ogelman}, {Starrfield},
  {Wichmann}, \& {Pfeffermann}}]{krautt96}
{Krautter}, J., {\"Ogelman}, H., {Starrfield}, S., {Wichmann}, R., \&
  {Pfeffermann}, E. 1996, ApJ, 456, 788

\bibitem[{{Krautter} \& {Williams}(1989)}]{1989ApJ...341..968K}
{Krautter}, J., \& {Williams}, R.~E. 1989, ApJ, 341, 968

\bibitem[{{Lanz} {et~al.}(2005){Lanz}, {Telis}, {Audard}, {Paerels},
  {Rasmussen}, \& {Hubeny}}]{lanz04}
{Lanz}, T., {Telis}, G.~A., {Audard}, M., {Paerels}, F., {Rasmussen}, A.~P., \&
  {Hubeny}, I. 2005, ApJ, 619, 517

\bibitem[{{MacDonald} {et~al.}(1985){MacDonald}, {Fujimoto}, \&
  {Truran}}]{macdonald85}
{MacDonald}, J., {Fujimoto}, M.~Y., \& {Truran}, J.~W. 1985, ApJ, 294, 263

\bibitem[{{Munari} {et~al.}(1996){Munari}, {Goranskij}, {Popova}, {Shugarov},
  {Tatarnikov}, {Yudin}, {Karitskaya}, {Kusakin}, {Zwitter}, {Lepardo},
  {Passuello}, {Sostero}, {Metlova}, \& {Shenavrin}}]{1996A&A...315..166M}
{Munari}, U., {Goranskij}, V.~P., {Popova}, A.~A., {Shugarov}, S.~Y.,
  {Tatarnikov}, A.~M., {Yudin}, B.~F., {Karitskaya}, E.~A., {Kusakin}, A.~V.,
  {Zwitter}, T., {Lepardo}, A., {Passuello}, R., {Sostero}, G., {Metlova},
  N.~V., \& {Shenavrin}, V.~I. 1996, A\&A, 315, 166

\bibitem[{{Ness} {et~al.}(2007{\natexlab{a}}){Ness}, {Schwarz}, {Retter},
  {Starrfield}, {Schmitt}, {Gehrels}, {Burrows}, \& {Osborne}}]{swnovae}
{Ness}, J.-U., {Schwarz}, G.~J., {Retter}, A., {Starrfield}, S., {Schmitt},
  J.~H.~M.~M., {Gehrels}, N., {Burrows}, D., \& {Osborne}, J.~P.
  2007{\natexlab{a}}, ApJ, 663, 505

\bibitem[{{Ness} {et~al.}(2007{\natexlab{b}}){Ness}, {Starrfield}, {Beardmore},
  {Bode}, {Drake}, {Evans}, {Gehrz}, {Goad}, {Gonzalez-Riestra}, {Hauschildt},
  {Krautter}, {O'Brien}, {Osborne}, {Page}, {Sch\"onrich}, \&
  {Woodward}}]{ness_rsoph}
{Ness}, J.-U., {Starrfield}, S., {Beardmore}, A., {Bode}, M.~F., {Drake},
  J.~J., {Evans}, A., {Gehrz}, R., {Goad}, M., {Gonzalez-Riestra}, R.,
  {Hauschildt}, P., {Krautter}, J., {O'Brien}, T.~J., {Osborne}, J.~P., {Page},
  K.~L., {Sch\"onrich}, R., \& {Woodward}, C. 2007{\natexlab{b}}, ApJ, 665,
  1334

\bibitem[{{Ness} {et~al.}(2003){Ness}, {Starrfield}, {Burwitz}, {Wichmann},
  {Hauschildt}, {Drake}, {Wagner}, {Bond}, {Krautter}, {Orio}, {Hernanz},
  {Gehrz}, {Woodward}, {Butt}, {Mukai}, {Balman}, \& {Truran}}]{v4743}
{Ness}, J.-U., {Starrfield}, S., {Burwitz}, V., {Wichmann}, R., {Hauschildt},
  P., {Drake}, J.~J., {Wagner}, R.~M., {Bond}, H.~E., {Krautter}, J., {Orio},
  M., {Hernanz}, M., {Gehrz}, R.~D., {Woodward}, C.~E., {Butt}, Y., {Mukai},
  K., {Balman}, S., \& {Truran}, J.~W. 2003, ApJL, 594, L127

\bibitem[{{Ness} {et~al.}(2007{\natexlab{c}}){Ness}, {Starrfield}, {Osborne},
  {Page}, \& {Schwarz}}]{ness_rrtel}
{Ness}, J.-U., {Starrfield}, S., {Osborne}, J., {Page}, K.~L., \& {Schwarz}, G.
  2007{\natexlab{c}}, Central Bureau Electronic Telegrams (2007).~ Edited by
  Green, D.~W.~E., submitted

\bibitem[{{Ness} {et~al.}(2006){Ness}, {Starrfield}, {Schwarz},
  {Vanlandingham}, {Wagner}, {Lyke}, {Woodward}, {Lynch}, \&
  {Krautter}}]{2006IAUC.8676....2N}
{Ness}, J.-U., {Starrfield}, S., {Schwarz}, G., {Vanlandingham}, K., {Wagner},
  R.~M., {Lyke}, J., {Woodward}, C.~E., {Lynch}, D.~K., \& {Krautter}, J. 2006,
  IAU Circ, 8676, 2

\bibitem[{{\"Ogelman} {et~al.}(1993){\"Ogelman}, {Orio}, {Krautter}, \&
  {Starrfield}}]{oegel93}
{\"Ogelman}, H., {Orio}, M., {Krautter}, J., \& {Starrfield}, S. 1993, Nature,
  361, 331

\bibitem[{{Orio}(2006)}]{orio06}
{Orio}, M. 2006, ApJ, 643, 844

\bibitem[{{Orio} {et~al.}(2001){Orio}, {Covington}, \& {{\"O}gelman}}]{orio01}
{Orio}, M., {Covington}, J., \& {{\"O}gelman}, H. 2001, A\&A, 373, 542

\bibitem[{{Petz} {et~al.}(2005){Petz}, {Hauschildt}, {Ness}, \&
  {Starrfield}}]{petz05}
{Petz}, A., {Hauschildt}, P.~H., {Ness}, J.-U., \& {Starrfield}, S. 2005, A\&A,
  431, 321

\bibitem[{{Pietsch} {et~al.}(2005){Pietsch}, {Fliri}, {Freyberg}, {Greiner},
  {Haberl}, {Riffeser}, \& {Sala}}]{pietsch05}
{Pietsch}, W., {Fliri}, J., {Freyberg}, M.~J., {Greiner}, J., {Haberl}, F.,
  {Riffeser}, A., \& {Sala}, G. 2005, A\&A, 442, 879

\bibitem[{{Pietsch} {et~al.}(2007){Pietsch}, {Haberl}, {Sala}, {Stiele},
  {Hornoch}, {Riffeser}, {Fliri}, {Bender}, {B{\"u}hler}, {Burwitz}, {Greiner},
  \& {Seitz}}]{pietsch06}
{Pietsch}, W., {Haberl}, F., {Sala}, G., {Stiele}, H., {Hornoch}, K.,
  {Riffeser}, A., {Fliri}, J., {Bender}, R., {B{\"u}hler}, S., {Burwitz}, V.,
  {Greiner}, J., \& {Seitz}, S. 2007, A\&A, 465, 375

\bibitem[{{Read} {et~al.}(2005){Read}, {Saxton}, {Esquej}, {Freyberg}, \&
  {Altieri}}]{xmmsurvey}
{Read}, A.~M., {Saxton}, R.~D., {Esquej}, M.~P., {Freyberg}, M.~J., \&
  {Altieri}, B. 2005, in 5 years of Science with XMM-Newton, ed. U.~G. {Briel},
  S.~{Sembay}, \& A.~{Read}, 137

\bibitem[{{Sala} \& {Hernanz}(2005)}]{sala06}
{Sala}, G., \& {Hernanz}, M. 2005, A\&A, 439, 1061

\bibitem[{{Schlegel} {et~al.}(1998){Schlegel}, {Finkbeiner}, \&
  {Davis}}]{schlegel98}
{Schlegel}, D.~J., {Finkbeiner}, D.~P., \& {Davis}, M. 1998, ApJ, 500, 525

\bibitem[{{Schwarz} {et~al.}(2007{\natexlab{a}}){Schwarz}, {Woodward},
  {Starrfield}, {Vanlandingham}, {Wagner}, {Ness}, \& {Helton}}]{swaas07}
{Schwarz}, G., {Woodward}, C., {Starrfield}, S., {Vanlandingham}, K., {Wagner},
  M., {Ness}, J., \& {Helton}, A. 2007{\natexlab{a}}, in American Astronomical
  Society Meeting Abstracts, Vol. 210, American Astronomical Society Meeting
  Abstracts, 04.04

\bibitem[{{Schwarz} {et~al.}(2001){Schwarz}, {Shore}, {Starrfield},
  {Hauschildt}, {Della Valle}, \& {Baron}}]{schwarz_lmc}
{Schwarz}, G.~J., {Shore}, S.~N., {Starrfield}, S., {Hauschildt}, P.~H., {Della
  Valle}, M., \& {Baron}, E. 2001, MNRAS, 320, 103

\bibitem[{{Schwarz} {et~al.}(2007{\natexlab{b}}){Schwarz}, {Woodward}, {Bode},
  {Evans}, {Eyres}, {Geballe}, {Gehrz}, {Greenhouse}, {Helton}, {Liller},
  {Lyke}, {Lynch}, {O'Brien}, {Rudy}, {Russell}, {Shore}, {Starrfield},
  {Temim}, {Truran}, {Venturini}, {Wagner}, {Williams}, \&
  {Zamanov}}]{schwarz07}
{Schwarz}, G.~J., {Woodward}, C.~E., {Bode}, M.~F., {Evans}, A., {Eyres},
  S.~P., {Geballe}, T.~R., {Gehrz}, R.~D., {Greenhouse}, M.~A., {Helton},
  L.~A., {Liller}, W., {Lyke}, J.~E., {Lynch}, D.~K., {O'Brien}, T.~J., {Rudy},
  R.~J., {Russell}, R.~W., {Shore}, S.~N., {Starrfield}, S.~G., {Temim}, T.,
  {Truran}, J.~W., {Venturini}, C.~C., {Wagner}, R.~M., {Williams}, R.~E., \&
  {Zamanov}, R. 2007{\natexlab{b}}, AJ, 134, 516

\bibitem[{{Shanley} {et~al.}(1995){Shanley}, {\"Ogelman}, {Gallagher}, {Orio},
  \& {Krautter}}]{shanley95}
{Shanley}, L., {\"Ogelman}, H., {Gallagher}, J.~S., {Orio}, M., \& {Krautter},
  J. 1995, ApJL, 438, L95

\bibitem[{{Starrfield}(1989)}]{st89}
{Starrfield}, S. 1989, in Classical Novae, ed. M.~Bode \& A.~Evans (Wiley, New
  York), 39

\bibitem[{{Starrfield} {et~al.}(2004){Starrfield}, {Timmes}, {Hix}, {Sion},
  {Sparks}, \& {Dwyer}}]{starrf04}
{Starrfield}, S., {Timmes}, F.~X., {Hix}, W.~R., {Sion}, E.~M., {Sparks},
  W.~M., \& {Dwyer}, S.~J. 2004, ApJL, 612, L53

\bibitem[{{Truran} \& {Livio}(1986)}]{truranlivio86}
{Truran}, J.~W., \& {Livio}, M. 1986, ApJ, 308, 721

\bibitem[{{van den Heuvel} {et~al.}(1992){van den Heuvel}, {Bhattacharya},
  {Nomoto}, \& {Rappaport}}]{heuvel}
{van den Heuvel}, E.~P.~J., {Bhattacharya}, D., {Nomoto}, K., \& {Rappaport},
  S.~A. 1992, A\&A, 262, 97

\end{thebibliography}

\end{document}